\documentclass[aps,prd,preprintnumbers,showpacs]{revtex4}
\usepackage[dvips]{graphicx}
\begin{document}

%
%

\eprint{Nisho-2-2014}
\title{Proton Mass Shift in Muonic Hydrogen Atom }
\author{Aiichi Iwazaki}
\affiliation{International Economics and Politics, Nishogakusha University,\\ 
6-16 3-bantyo Chiyoda Tokyo 102-8336, Japan.}   
\date{August. 11, 2014}
\begin{abstract}
We show that the value of the proton mass depends on each bound state of muonic or electronic
hydrogen atom. The charged particle bound to the proton 
produces magnetic field inside the proton. This
makes a change to the amount of chiral condensate inside the proton.
The change gives rise to the shift in the value of the proton mass. 
Numerically, the shift in the $2S$ state of the muonic hydrogen atom can be of the order of $0.1$ meV.
The effect may solve the puzzle of the proton radius.
\end{abstract}
\hspace*{0.3cm}
\hspace*{1cm}

\maketitle

\section{introduction}
The proton mass would be much smaller than $1$ GeV
if the chiral symmetry was not spontaneously broken in QCD.
It would take a value of several MeV. 
However the real mass is about $1$ GeV which results from 
the spontaneous breaking of the chiral symmetry. The breaking generates the chiral
condensate $\langle \bar{q}q \rangle\simeq (-240\mbox{MeV})^3$, which leads to the
large proton mass. 
It has been discussed\cite{chiral,chiral2} that the chiral condensate
grows with external magnetic field. Since the proton has the magnetic moment,
it has an intrinsic magnetic field inside the proton itself. Thus,
the amount of the chiral condensate inside the proton is larger than 
that of the condensate in vacuum.
Furthermore, an electron or a muon bound to the proton
produces a magnetic field inside the proton, which makes a change to the amount of
the chiral condensate inside the proton. 
Although the change is much small,
it causes the shift in the value of the proton mass.

In this paper we show that the magnetic field produced by a muon or an electron bound to the proton
gives rise to the measurable shift in the value of the proton mass. 
The shift in the muonic hydrogen atom is much larger than that in the
electronic hydrogen atom. This is because the Bohr radius in the muonic hydrogen atom 
is much smaller than the Bohr radius in the electronic hydrogen atom.
Furthermore, the amount of the shift depends on each
bound state. As the charged particle is closer to the proton, 
it produces stronger magnetic field
inside the proton. 
Thus, the shift is larger in S states than that in the other
states with higher orbital
angular momentum $l \ge 1$. 
We point out that the recent puzzle\cite{puzzle,puzzle2} of the proton radius 
may be solved by taking account of the proton mass shift
in the muonic hydrogen atom.

The condensate inside the proton is exposed to two types of magnetic fields.
The one is intrinsic magnetic field $B_i$ and the other one is
external magnetic field $\delta B$. 
The intrinsic magnetic field 
is its own magnetic field inside the proton because the proton has a magnetic moment.
The external magnetic field is produced by an electron or muon bound to the proton. 
The strength of the external magnetic field depends on each bound state of the particles. 
As naively expected, the external magnetic field is much smaller than the intrinsic one.
These magnetic fields change the amount of the chiral condensate so that 
they change the value of the proton mass.
Although the proton mass shift caused by the external magnetic field is very small,
we can measure the proton mass shift by observing emission spectrum of muonic
or electronic hydrogen atom.  






\vspace{0.2cm}
In order to find the mass shift caused by the magnetic field of the bound muon or electron,
we first represent the proton mass $M_P$ in term of the chiral condensate.
We assume that the Ioffe's formula\cite{Ioffe} holds under background magnetic field 
$\vec{B}=\vec{B}_i+\delta\vec{B}$ 

\begin{equation}
\label{Ioffe}
M_P=\frac{-4\pi^2 \langle \bar{q}q \rangle _{B}}{\Lambda_{QCD}^2}
\end{equation}
where $\Lambda_{QCD}$ denotes a constant characterizing QCD scale; $\Lambda_{QCD}\simeq 1$GeV.
We use the units of $c=1$ and $\hbar=h/2\pi=1$.
The formula is not precisely but approximately derived by using QCD sum rule.
The question how the proton mass depends on the condensate is still controversial.
Hence, our estimation of the proton mass shift with the use of the formula in eq(\ref{Ioffe}) 
should be taken as being not seriously quantitative,
only approximately quantitative.
The point in the formula is 
that the proton mass depends on the magnetic field only through the chiral condensate.
Thus, the effect of the small external magnetic field $\delta B$ added to $B_i$ 
on the proton mass is given such that

\begin{equation}
\label{sh}
\delta M_{\delta B}=M_P\frac{\delta \langle \bar{q}q \rangle _{B_i}}{\langle \bar{q}q \rangle _{B_i}}, \quad
\delta \langle \bar{q}q \rangle _{B_i}\equiv \langle \bar{q}q 
\rangle _{B_i+\delta B}-\langle \bar{q}q \rangle _{B_i}.
\end{equation}  

We note that in the case that $M_P\propto (-\langle \bar{q}q \rangle_{B})^{1/3}$ 
as anticipated from dimensional analysis,
the above formula is rewritten as 
$\delta M_{\delta B}=M_P\delta \langle \bar{q}q \rangle _{B_i}/3\langle \bar{q}q \rangle _{B_i}$
The difference between these two formulas is the factor $1/3$.
Hence, owing to the ambiguity of the dependence of the proton mass on the chiral condensate,
subsequent our result of the proton mass shift has an error of the order of a numerical factor.

In addition to the Ioffe's formula,
we assume that the intrinsic magnetic field $\vec{B}_i$ inside the proton
is proportional to the proton magnetic moment $\vec{\mu}_p=g_p\vec{S_p} e/2M_P$,
i.e. $\vec{B_i}\propto \vec{\mu}_p$, 
where $\vec{S}_p$ denotes the proton spin and $g_p\simeq 5.6$.
In principle,
we can precisely determine the value $B_i$ in the lattice gauge theory, although
it has not yet performed. Instead, we may obtain the approximate value $B_i$ such that
$B_i\simeq g_pe/8\pi r_p^3M_p\sim (29\mbox{MeV})^2$
with the proton radius $r_p=0.88$fm,
assuming that
a circular current $I$ with
its radius $r_p$ has the magnetic moment $\mu_p=\pi r_p^2I$ ( $=g_pe/4M_P$ ) and  
generates a magnetic field $B_i=I/2r_p$ at the center of the 
circle. But,  
in this paper we take it as one of unknown parameters.
The magnetic field $\delta \vec{B}$ produced by the electron or muon bound to the proton is
added to this intrinsic magnetic field.

Secondly in order to calculate the mass shift in eq(\ref{sh}), we need an explicit formula representing
the dependence of the chiral condensate on the background magnetic field. It has been
shown\cite{chiral2} that the condensate depends on $B$ such that

\begin{equation}
\label{cond}
\langle \bar{q}q \rangle _{B}=\langle \bar{q}q \rangle _{B=0}
\Big(1+\frac{(eB)^2}{96\pi^2F_{\pi}^2m_{\pi}^2}\Big),
\end{equation}
in the limit $eB\ll m_{\pi}^2$, where $F_{\pi}\simeq 90$MeV ( $m_{\pi}$ ) denotes pion decay constant 
( pion mass ). The formula has been explicitly derived 
by using chiral perturbation theory\cite{scherer}, i.e., low energy model of
hadrons. It has been partially confirmed in lattice gauge theories\cite{lattice}
where the value of the magnetic field
is limited to be in the range $(180\mbox{MeV})^2 <eB<(700\mbox{MeV})^2$ and 
the pion mass is given by $m_{\pi}\simeq 200$MeV.
That is, the quadratic dependence on $B$
has been confirmed although $B$ is limited to the range. But,
the coefficient of the term $\propto B^2$ has been found  
to be smaller in the lattice gauge theories compared with the theoretical
prediction in eq(\ref{cond}).
Thus, 
the formula in eq(\ref{cond}) is not still confirmed. 
Hence, we take the value $F_{\pi}^2m_{\pi}^2$ or $F_{\pi}^2$ 
as one of unknown parameters, although we use the formula in eq(\ref{cond}) even in the case of the small
magnetic field $B\sim (29\mbox{MeV})^2\ll m_{\pi}^2$.   

Using the formula, we obtain the mass shift caused by a bound muon or electron 

\begin{equation}
\label{shift}
\delta M_{\delta B}=M_P\frac{e\vec{B_i}\cdot e\delta\vec{B}}{48\pi^2F_{\pi}^2m_{\pi}^2}
\end{equation}

The magnetic field $\delta \vec{B}$ produced by the bound muon or electron is  given by

\begin{equation}
\delta\vec{B}=\mu_{B,\mu,e}\frac{\vec{l}_{\mu,e}}{4\pi r^3}+
\frac{1}{4\pi r^3}\Big(\frac{3(\vec{\mu}_{\mu,e}\cdot\vec{r})\vec{r}}{r^2}-\vec{\mu}_{\mu,e}\Big)+
\frac{2\vec{\mu}_{\mu,e}}{3}\delta^3(r),
\end{equation}  
where
$\vec{\mu}_{\mu,e}$ ( $=-g_{\mu,e}\mu_{B,\mu,e}\vec{S}_{\mu,e}$ ) denotes the magnetic moment 
of the muon or electron with $\mu_{B,\mu,e}\equiv e/2m_{\mu,e}$ and $g_{\mu,e}\simeq 2$;
$\vec{S}_{\mu,e}$ denotes the spin of the muon or electron.
The first term represents a contribution of  the orbital angular momentum $\vec{l}$, and
the second and third terms represent contributions of the magnetic moment $\vec{\mu}_{\mu,e}$.
The muon or electron is located  at $\vec{r}=0$ in the coordinate.    

Therefore, taking the quantum average with the use of the wavefunction of the muon or electron, 
the proton mass shift in eq( $\ref{shift}$ ) is given by

\begin{equation}
\label{shift2}
\langle \delta M_{\delta B}\rangle=\frac{eM_P}{48\pi^2F_{\pi}^2m_{\pi}^2}\Bigg(
\langle \frac{\mu_{B,\mu,e}e\vec{B_i}\cdot\vec{l}_{\mu,e}}{4\pi r^3}\rangle 
+\langle \frac{1}{4\pi r^3}\Big(\frac{3(\vec{\mu}_{\mu,e}\cdot\vec{r})(e\vec{B_i}\cdot\vec{r})}{r^2}
-e\vec{B_i}\cdot\vec{\mu}_{\mu,e}\Big)\rangle+
\langle\frac{2e\vec{B_i}\cdot \vec{\mu}_{\mu,e}}{3}\delta^3(r)\rangle \Bigg ),
\end{equation} 
where the coordinate $\vec{r}$ denotes the location of the bound particles;
the proton is located at $\vec{r}=0$.

Since the first and second terms vanish for the S states with $l=0$,
we can easily obtain the mass shift in the S states with the principal quantum number $n$,

\begin{equation}
\langle \delta M_{\delta B}\rangle=
\frac{eM_P}{48\pi^2F_{\pi}^2m_{\pi}^2}\frac{-2eB_i \mu_{\mu,e} |\Psi_n^{l=0}(\vec{r}=0)|^2}{3}
\Big(F(F+1)-\frac{3}{2}\Big)
\end{equation}
with $\mu_{\mu,e}\equiv g_{\mu,e}e/2m_{\mu,e}$ 
and total spin $F=1$ or $F=0$ of the proton and muon ( electron ),
where we have set $\vec{B}_i\equiv 2B_i\vec{S}_p$.
We have used nonrelativistic hydrogen wave functions $\Psi_n^{l=0}(\vec{r})$
with angular momentum $l=0$, principal quantum number $n$ and
reduced mass $\bar{m}_{\mu,e}\equiv m_{\mu,e}M_P/(M_P+m_{\mu,e})$.

For example, in the state $2S_{1/2}^{F=1}$ ( n=2 ) of the muonic hydrogen atom 

\begin{equation}
\label{2S}
\langle \delta M_{\delta B}(\mbox{muon})\rangle_{2S}=
\frac{eM_P}{48\pi^2F_{\pi}^2m_{\pi}^2}\frac{-eB_i \mu_{\mu}(\alpha \bar{m}_{\mu})^3}{24\pi}=
\frac{-\alpha^4 \bar{m}_{\mu}^3 M_P}{288\pi^2m_{\pi}^2m_{\mu}}\frac{eB_i}{F_{\pi}^2}
\simeq -0.39\,\,\frac{eB_i}{F_{\pi}^2}\,\mbox{meV},
\end{equation}
with $\alpha\equiv e^2/4\pi \simeq 1/137$,
where the values $M_P=938$MeV, $m_{\mu}=106$MeV and $m_{\pi}=140$MeV are used.
The factor $eB_i/F_{\pi}^2$ is an unknown parameter.
The negative value $\langle \delta M_{\delta B}\rangle_{2S}$ 
comes from the fact that the magnetic moment of the proton is
anti-parallel to the magnetic moment of the muon in the state $2S_{1/2}^{F=1}$.
Similarly, we may estimate the mass shifts in the other S states
by the explicit use of their wave functions.
We should notice that as naively expected, 
the external magnetic field $e\delta B=\alpha^4\bar{m}_{\mu}^3/6m_{\mu}\sim (10^{-3}\mbox{MeV})^2$ 
extracted from the above formula 
is much smaller than the intrinsic magnetic
field $eB_i\sim (17\mbox{MeV})^2$.

\vspace{0.1cm}
Furthermore, in order to estimate the mass shifts in the states with higher angular momenta $l\ge 1$
we only need to estimate the first and second terms in eq(\ref{shift2}) since
the third term vanishes. For example, we shall estimate the proton mass shift in the state
 $2P_{3/2}^{F=2}$ of the muonic hydrogen atom where
the $\vec{F}$ denotes the total angular momentum of the proton and the muon; 
$\vec{F}=\vec{S_p}+\vec{S}_{\mu}+\vec{l}_{\mu}$. 
In the calculation we take the average of $\langle \vec{S_p}\cdot \vec{l}_{\mu} \rangle$
and $\langle \frac{3(\vec{S}_{\mu}\cdot\vec{r})(\vec{S}_p\cdot\vec{r})}{r^2}-\vec{S}_{\mu}\cdot\vec{S}_p\rangle $
over the spin and the angular coordinates

\begin{equation}
\langle \vec{S_p}\cdot \vec{l}_{\mu} \rangle=\frac{1}{2}, \quad
\langle \frac{3(\vec{S}_{\mu}\cdot\vec{r})(\vec{S}_p\cdot\vec{r})}{r^2}-\vec{S}_{\mu}\cdot\vec{S}_p\rangle
=-\frac{1}{10}.
\end{equation}
Finally taking the average of $1/r^3$ over the radial coordinate $r$ with the explicit use 
of the wave function $\Psi_{n=2}^{l=1}(\vec{r})$, we obtain

\begin{equation}
\label{2P}
\langle \delta M_{\delta B}(\mbox{muon})\rangle_{2P}=
\frac{7\alpha^4 \bar{m}_{\mu}^3 M_P}{11520\pi^2m_{\pi}^2m_{\mu}}\frac{eB_i}{F_{\pi}^2}
\simeq 0.068\,\frac{eB_i}{F_{\pi}^2}\,\mbox{meV}.
\end{equation}
where the identical values $M_P,\,m_{\mu}$ and $m_{\pi}$ to those in eq(\ref{2S}) was used.  
As we have expected, 
we find that the mass shift in the S state is much larger than that in the P state. 
In this way
we can evaluate the proton mass shift in any bound states of the muon or electron
by using the formula in eq(\ref{shift2}).
We should notice that the mass shift is caused by non-perturbative effect,
since the nonvanishing chiral condensate arises owing to the spontaneous 
breaking of the chiral symmetry. 
Thus, the proton mass shift 
has not been previously found in any perturbative
analyses of the muonic hydrogen spectra.

\vspace{0.2cm}
Now, we point out a possible solution for the puzzle\cite{puzzle,puzzle2} of the proton radius.
The puzzle arises from the discrepancy between the theoretical prediction
of the photon energy and the observed one of the muonic transition in the muonic hydrogen atom,

\begin{equation}
\label{dis}
\Big( E(2P_{3/2}^{F=2})-E(2S_{1/2}^{F=1})\Big)_{\rm obs}-
\Big( E(2P_{3/2}^{F=2})-E(2S_{1/2}^{F=1})\Big)_{\rm th}=0.31\,\mbox{meV}
\end{equation} 
where a muon emits a photon in the transition from the state $2P_{3/2}^{F=2}$ with the energy
$E(2P_{3/2}^{F=2})$ to the state $2S_{1/2}^{F=1}$ with the lower energy $E(2S_{1/2}^{F=1})$.
The CODATA-2010 value of the proton radius $r_p\simeq 0.88$fm 
was used in the theoretical prediction, which has been
obtained \cite{codata} from the analysis of the electronic hydrogen
spectra and 
the experiment of electron-proton scattering. 
On the other hand, the theoretical prediction
agrees with the observed one when one uses the smaller radius $ r_p\simeq 0.84$fm.
The theoretical prediction of the spectrum in the muonic hydrogen has been
examined\cite{puzzle2} in detail, but there have been no rooms 
found for the improvement except for the change of the proton radius.
This is the puzzle of the proton radius.

Our proposal for the solution of the puzzle is that the discrepancy in eq(\ref{dis}) may be solved by
taking account of the proton mass shift caused by the magnetic field of the bound muon.
Namely, when the proton mass shift is taken into account, 
the above formula is improved 
such that 

\begin{equation}
\label{dis2}
\Big( E(2P_{3/2}^{F=2})-E(2S_{1/2}^{F=1})\Big)_{\rm obs}-
\Big( E(2P_{3/2}^{F=2})-E(2S_{1/2}^{F=1})\Big)_{\rm th}=
0.31\,\mbox{meV}-0.46\frac{eB_i}{F_{\pi}^2}\mbox{meV},
\end{equation}
with the proton radius $r_p\simeq 0.88$fm being used. In order to determine the parameter 
$eB_i/F_{\pi}^2$, we need to obtain 
the intrinsic magnetic field $B_i$
and the coefficient $F_{\pi}^2m_{\pi}^2$ in eq(\ref{cond}), both of which have not yet 
obtained in lattice gauge theories.
When we tentatively adopt the approximate value $B_i\sim (30\mbox{MeV})^2$
and use the pion decay constant $F_{\pi}\simeq 90$MeV, we obtain $eB_i/F_{\pi}^2\sim 0.04$.
Hence, the mass shift is slightly too small to solve the puzzle. 
But we should remember that there are several ambiguities of the physical parameters
$B_i$, $F_{\pi}^2m_{\pi}^2$ and of the dependence of the proton mass on
the chiral condensate in eq(\ref{Ioffe}). 
In addition to the ambiguities, we need to reexamine the analysis
of the electronic hydrogen spectra to obtain the proton radius $r_p$
by taking into account the proton mass shift. 
Then, we find that when we 
include the effect of the proton mass shift in the analysis of the electronic hydrogen spectra, we obtain 
the proton radius $r_p$ smaller than the value $0.88$fm which was obtained without
the inclusion of the effect of the mass shift. 
The smaller proton radius leads to the discrepancy with smaller amount
than $0.31$meV in eq(\ref{dis}).
Then, the small parameter $eB_i/F_{\pi}^2\sim 0.04$ obtained approximately
would be suitable for the discrepancy. 

The proton mass shift is fairly small 
in the case of the electronic hydrogen atom because the Bohr radius of the electron is $200$
times larger than that of the muon. Actually,
the shift can be easily obtained by replacing the muon mass by the electron mass
in eq(\ref{2S}) and eq(\ref{2P}),

\begin{eqnarray}
\langle \delta M_{\delta B}(\mbox{electron})\rangle_{2S}&=&
\frac{\alpha^4 \bar{m}_e^3 M}{288\pi^2m_{\pi}^2m_e}\frac{eB_i}{F_{\pi}^2}
\simeq -1.2\times 10^{-5}\frac{eB_i}{F_{\pi}^2}\,\mbox{meV}\nonumber \\
\langle \delta M_{\delta B}(\mbox{electron})\rangle_{2P}&=&
\frac{7\alpha^4 \bar{m}_e
^3 M}{11520\pi^2m_{\pi}^2m_e}\frac{eB_i}{F_{\pi}^2}\simeq 2.2\times 10^{-6}\frac{eB_i}{F_{\pi}^2}\,\mbox{meV}
\end{eqnarray}
where the value $m_e=0.51$MeV is used.
These values are very small, but they are measurable because the accuracy of
the observed spectrum is smaller than the values.

\vspace{0.1cm}
To summarize, we have shown that the value of the proton mass in atoms
is different from that in vacuum. This is 
because the chiral condensate inside the proton 
is influenced by the magnetic field of bound
electrons or muons.
We have calculated the shifts in the value of the proton mass
in the muonic or electronic hydrogen atom. In particular we have explicitly shown
the mass shifts in the bound states $2S_{1/2}^{F=1}$ and $2P_{3/2}^{F=2}$ of the muonic and
electronic hydrogen atom.
We have found that the mass shifts give rise to a possible solution 
for the puzzle of the proton radius.

Taking account of the proton mass shift, 
the values of the fine structure constant $\alpha=e^2/4\pi$, Rydberg constant 
$R_{\infty}\equiv\alpha^2m_e/4\pi$
and the proton radius $r_p$ have to be 
reexamined in the analysis of the electronic hydrogen spectra. 

\vspace*{0.2cm}
We would like to express our thanks to Prof. K. Fukushima at University of tokyo 
for his useful discussion and comments.


\end{document}